\documentclass[useAMS]{mn2e}  
\usepackage{graphicx}

\title[The incomplete Paschen-Back effect in Ap stars]
      {Modelling the incomplete Paschen-Back effect in the 
      spectra of magnetic Ap stars}

\author[M.J.~Stift, F.~Leone and E.~Landi~Degl'Innocenti]
       {M.J.~Stift$^1$,
        F.~Leone$^2$,
        and E.~Landi~Degl'Innocenti$^3$\\
        $^1$Institut f{\"u}r Astronomie (IfA), Universit{\"a}t Wien,
            T{\"u}rkenschanzstrasse 17, A-1180 Wien, Austria\\
        $^2$Universit{\`a} di Catania, Dipartimento di Fisica e
            Astronomia -- Sezione Astrofisica, Via~S.~Sofia~78, I-95123
            Catania, Italy\\
        $^3$Universit{\`a} degli Studi di Firenze, Dipartimento di Astronomia
            e Scienza dello Spazio, Largo Enrico Fermi 2, I-50125 Firenze, Italy }
\begin{document}

\date{Accepted 2008}

\pagerange{\pageref{firstpage}--\pageref{lastpage}} \pubyear{2008}

\maketitle

\label{firstpage}

   \begin{abstract}
     {We present first results of a systematic investigation of the incomplete
      Paschen-Back effect in magnetic Ap stars. A short overview of the theory 
      is followed by a demonstration of how level splittings and component 
      strengths change with magnetic field strength for some lines of special 
      astrophysical interest. Requirements are set out for a code which allows 
      the calculation of full Stokes spectra in the Paschen-Back regime and 
      the behaviour of Stokes $I$ and $V$ profiles of transitions in the
      multiplet 74 of Fe\,{\sc ii} is discussed in some detail. It is shown that
      the incomplete Paschen-Back effect can lead to noticeable line shifts
      which strongly depend on total multiplet strength, magnetic field strength
      and field direction. Ghost components (which violate the normal selection
      rule on $J$) show up in strong magnetic fields but are probably
      unobservable. Finally it is shown that measurements of the integrated 
      magnetic field modulus $H_s$ are not adversely affected by the 
      Paschen-Back effect, and that there is a potential problem in (magnetic) 
      Doppler mapping if lines in the Paschen-Back regime are treated in the 
      Zeeman approximation.} 
   \end{abstract}


\section{Introduction}
\label{sec:intro}

Ever since Pieter Zeeman (1897) discovered the splitting of spectral lines
in a magnetic field, astrophysicists have tried to take advantage of the
diagnostic capabilities of this effect. Thanks to the Zeeman effect, G.E. Hale
(1908) was able to demonstrate the presence of strong magnetic fields in 
sunspots and over the years countless Zeeman observations (of ever increasing 
precision) of lines originating in all parts of the solar atmosphere have 
provided deep insights into the physics of the outer layers of the sun. The 
discovery by Friedrich Paschen and Ernst Back (1921) that in strong fields 
some transitions change from the anomalous Zeeman effect to the normal Zeeman 
effect, has not resulted in any revision of the traditional interpretation of
the observations, because even in the strongest solar magnetic fields (about 
0.4\,T) only very few lines are noticeably affected. Not until almost 50 years 
later was the question of the Paschen-Back (PB) effect in the solar spectrum first 
addressed by Engvold et al. (1970). Similarly, after the discovery by 
H.W. Babcock, starting 1947, of very strong magnetic fields in some upper-main 
sequence chemically peculiar stars (Ap stars), no serious thought was given to
the Paschen-Back effect. Although it is clear that in a 3.4\,T field like 
that of HD\,215441 (Babcock 1960), many spectral lines would exhibit the
transition from the Zeeman to the PB regime (which we shall call the incomplete 
or partial PB effect), the first papers on the Paschen-Back effect in a stellar 
context appeared only much later (Kemic 1975; Stift 1977).

Recently, interest in the Paschen-Back effect has revived, mainly in the
solar context and for molecular lines. As to atomic transitions, let us
note that the He\,{\sc i}\,10830\,{\AA} triplet has received special attention 
(Socas Navarro et al. 2004), as has for example the Mn\,{\sc i}\,15262.7\,{\AA} 
line (Asensio Ramos et al. 2007). In the stellar context, the work of Mathys 
(1990) constitutes a particularly detailed modelling attempt and discussion of 
the incomplete PB effect in the Fe\,{\sc ii} lines at $\lambda$\,6147.7 
and $\lambda$\,6149.2. The PB effect on hyperfine structure and its 
observational consequences have been investigated by Landolfi et al. (2001). 
It should be kept in mind that these results have generally been obtained in the 
Milne-Eddington approximation, and that heavy blending as often found in Ap stars 
could not be taken into account.

\begin{figure}
\includegraphics[width=84mm]{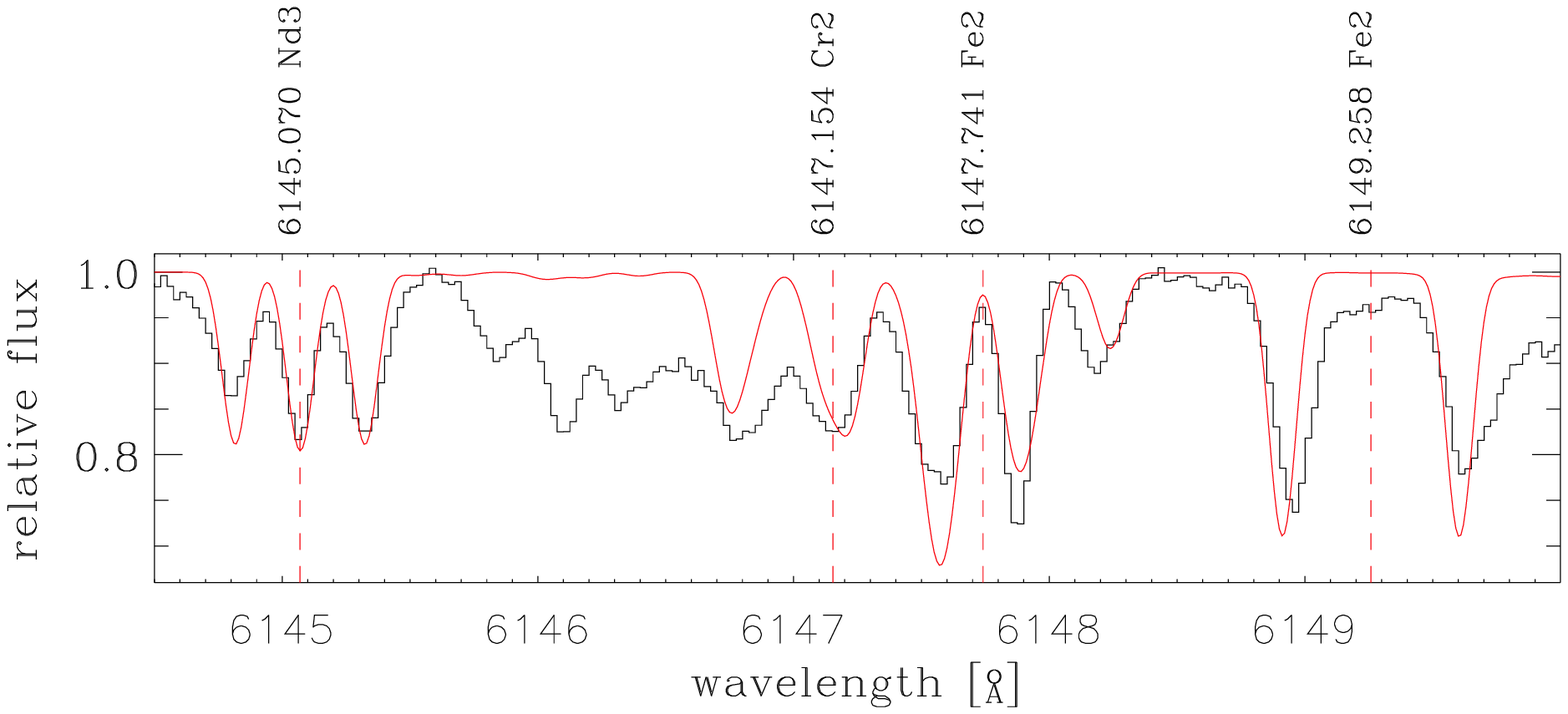}
\includegraphics[width=84mm]{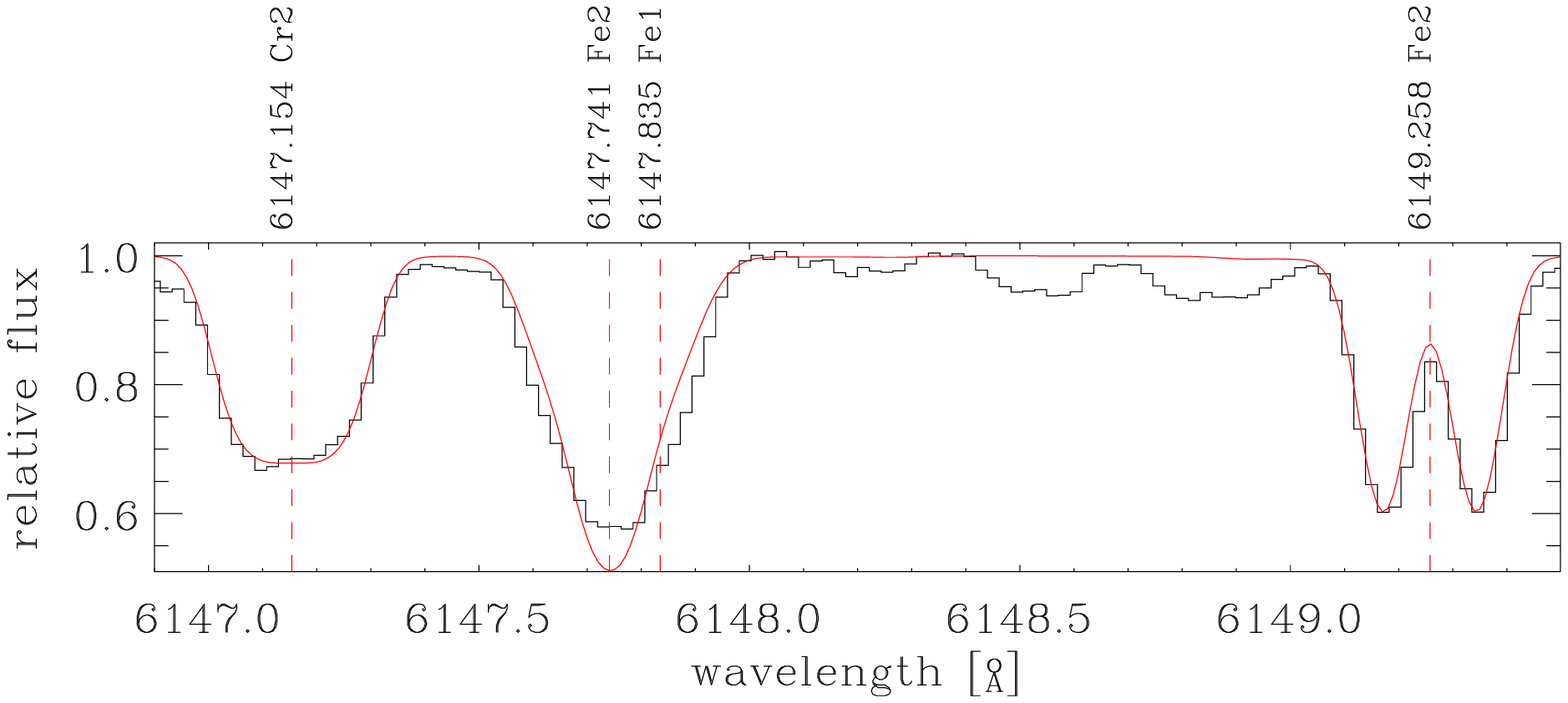}
\caption{
RV-corrected Stokes\,$I$ spectra in the vicinity of the Fe\,{\sc ii} lines at 6147
and 6149\,{\AA} of the stars HD\,66318 (top) and HD\,188041 (bottom). The magnetic
field modulus is 1.5\,T for the former star, 0.36\,T for the latter. The dashed 
red lines give the zero field positions of the lines of Cr, Fe and Nd. The full 
red lines are plotted for the only purpose of illustrating the line shift in 
$\lambda\,6149$; they result from a straightforward spectral synthesis, assuming 
normal Zeeman splitting. The RV has been determined from the magnetic null lines
($g_{\rm eff} = 0.0$) of Fe\,{\sc i} at $\lambda\,5434.52$ and $\lambda\,5576.09$.} 
\label{fig:shift}
\end{figure}

Our interest in the Paschen-Back effect has been stimulated by the difficulties
we encountered when trying to match high resolution spectral observations of 
strongly magnetic Ap stars with synthetic spectra (see Fig.\,\ref{fig:shift}).
The Zeeman doublet of Fe\,{\sc ii} at 6149\,{\AA} in particular not only 
displays non-symmetric relative intensities of the components -- already noted 
by Mathys (1990) -- but it also proves impossible to model the velocity 
shift (relative to the reference RV determined from magnetic null lines of iron) 
within the Zeeman regime. From Fig.\,\ref{fig:shift} it emerges that even at the 
very moderate field strength of 0.36\,T this shift is already observable.

Assuming that the Paschen-Back effect is responsible for both asymmetries and
shifts, the question naturally arises whether the measurement of the 
magnetic field modulus $H_s$ (the absolute field strength integrated over the 
visible hemisphere), which is based on the interpretation of the observed 
splitting of the $\lambda\,6149$ line in terms of simple Zeeman splitting, may
be systematically affected by the PB effect. This has not been studied so far, 
and we are also not aware of any discussion in the literature of the observed
shift of the $\lambda\,6149$ line.  

In the past, neither computers nor codes were powerful enough to embark on
any ambitious program of modelling stellar Stokes profiles in the incomplete 
PB regime. Now that multi-processor and/or multi-core architectures
have become affordable, COSSAM (Stift 2000) -- a state-of-the-art  object-oriented 
and fully parallel Stokes code -- has been modified in such a way as to allow 
the calculation for realistic stellar atmospheres of a multiplet in the 
incomplete PB regime, with full blending from the remaining spectral 
lines which are assumed to display classical anomalous Zeeman patterns. 
These tools make it possible to systematically explore how the observed Stokes
profiles are affected by the partial PB effect, to have a close look 
at the diagnostic content of Paschen-Back lines, and to model in detail selected 
spectral intervals in Ap stars with very strong fields. First important results 
are presented in this paper.

\section{The incomplete Paschen-Back effect}
\label{sec:paschen}

Named after the two German physicists Friedrich Paschen and Ernst Back, this 
effect generalises to magnetic fields of arbitrary strengths the better known
Zeeman effect. The effect was discovered in the laboratory in various 
multiplets (Paschen \& Back 1921), including the 
$4s \, 4d \, ^3D \rightarrow 4s \, 4p \, ^3P$ multiplet of Zn\,{\sc i} that was later
studied in major detail by van Geel (1928). The Paschen-Back effect has been 
successfully interpreted within the framework of quantum mechanics, and
nowadays this interpretation can be found in classical textbooks of atomic 
and/or molecular spectroscopy (see e.g. Condon \& Shortley 1935). 

In order to recall the basic physical facts, let us consider a term of an atom, 
characterised by the quantum numbers $L$ and $S$. The term is composed of
$2S+1$ (or $2L+1$ if $L < S$) $J$-levels whose energy separation is due to
the spin-orbit interaction. When a weak magnetic field is present, each 
$J$-level splits into $2J+1$ magnetic sublevels that can be identified by
the further quantum number $M$. In as far as the magnetic splitting is much lower 
than the fine-structure separation between different $J$-levels, such splitting 
turns out to be proportional to the magnetic field, and the atom is said to be 
in the Zeeman effect regime. However, when the magnetic field starts to be
comparable, or even larger than the fine-structure separation between 
$J$-levels, the linearity property is lost and the atom enters the 
regime of the (incomplete) Paschen-Back effect. Here the situation is
more complicated because the quantum number $J$ is no longer a good quantum
number and the atomic levels can thus be characterised only by the magnetic
quantum number $M$, not by $J$.

The energy values of the levels (the eigenvalues) and the corresponding 
eigenvectors can be obtained by diagonalisation of a set of matrices. 
The details of this procedure and the relevant equations are
described in full detail in Sect. 3.4 of the monograph by Landi Degl'Innocenti 
\& Landolfi (2004). In the same book one can also find the expressions for the 
splitting and the strengths of the various components ($\sigma_{\rm red}, \,
\pi, \; {\rm and} \; \sigma_{\rm blue}$) that arise in the transition between
a lower and an upper term, both in the Paschen-Back regime. The remarkable fact
is the appearance of some extra-components that can be referred to as 
"satellite" or "ghost" components. These components are strictly forbidden
in the absence of a magnetic field, and have negligible strengths in the
Zeeman effect regime, because of the selection rule 
$\Delta J = \pm \, 1, \, 0$, $0 \rightarrow \!\!\!\!\!\!\! / \;\;\; 0$. 
Referring for instance to the above-mentioned multiplet of Zn\,{\sc i}, the 
multiplet contains 6 lines in the Zeeman regime ($J_{\rm low} = 0 \rightarrow 
J_{\rm up} = 1, \, 
1 \rightarrow 1, \, 1 \rightarrow 2, \, 2 \rightarrow 1,  \ 2 \rightarrow 2,
\, 2\rightarrow 3$), but the number of lines increases to 9 in the Paschen-Back
regime. At the same time, the number of magnetic components increases
from 54 (18 $\sigma_{\rm red}$, 18 $\pi$, and 18 $\sigma_{\rm blue}$) in the
Zeeman regime to 71 (23 $\sigma_{\rm red}$, 25 $\pi$, and 23 
$\sigma_{\rm blue}$) in the Paschen-Back regime. 

The fact that strong 
magnetic fields are capable of bringing out lines which violate the ordinary
selection rule on $J$ was indeed the most important hint which led to the 
original discovery of the Paschen-Back effect. Examples of this behaviour
and further properties concerning the strengths and the splittings of the 
components are described in the following sections of the present paper. 

\begin{figure}
\includegraphics[width=84mm]{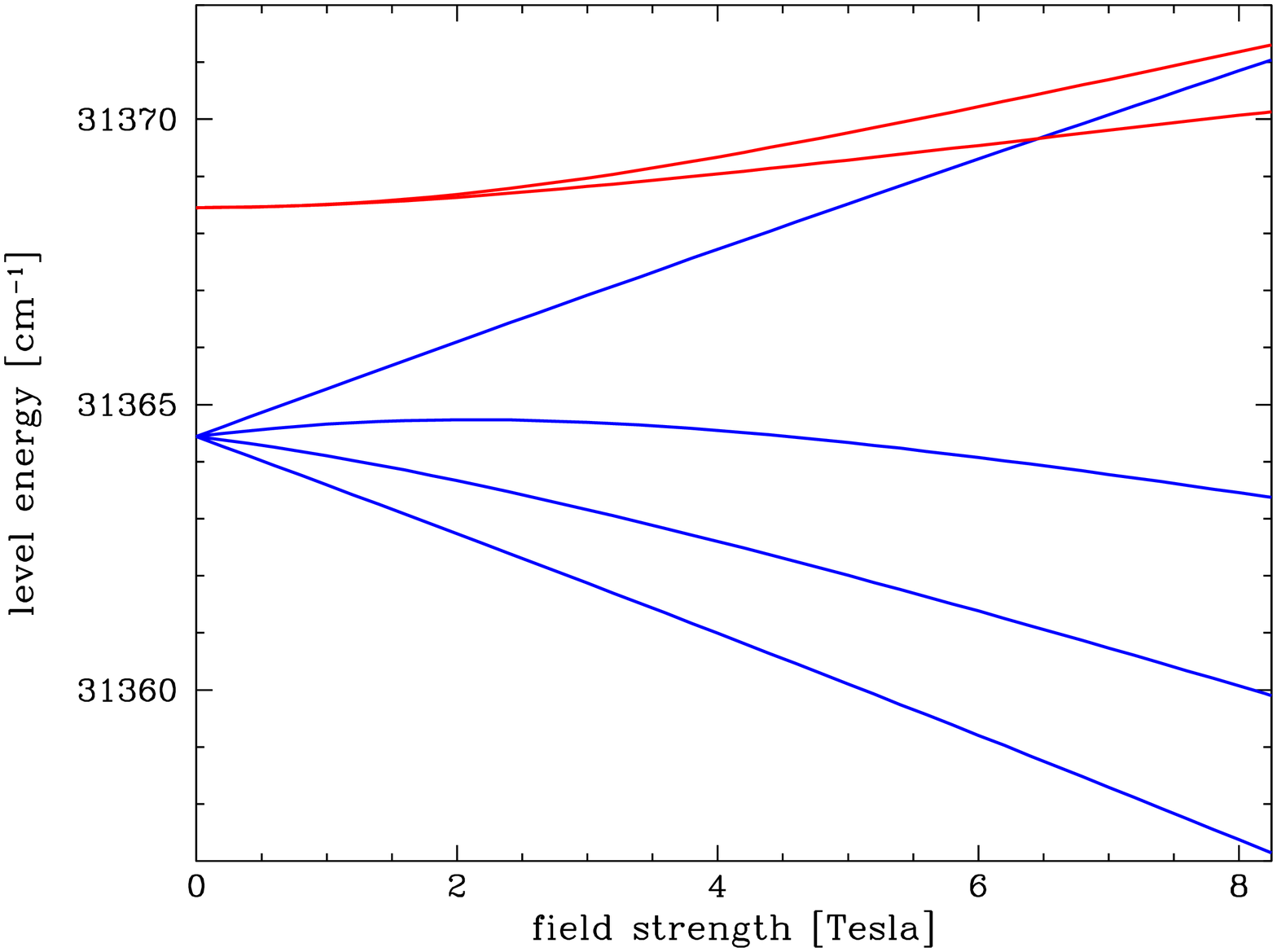}
\caption{
Magnetic splittings of the lowermost 2 levels of the ${^4}D$ term
of Fe\,{\sc ii} multiplet 74 as a function of field strength.}
\label{fig:6147_49_split}
\end{figure}

\section{Energy level splittings and relative component intensities}
\label{sec:energy}

The theory of the incomplete Paschen-Back effect reveals that both 
the splittings of the energy levels in a multiplet and the relative 
intensities of the subcomponents can change in a non-linear way with 
the magnetic field strength. Whether or not this happens at field
strengths typical for Ap stars (0.1 - 3.5\,T) depends on the detailed
fine structure splittings of the terms involved. It is useful to
recall that a field strength of 1\,T corresponds to a splitting of 
$0.47\,g\,M\,{\rm cm}^{-1}$, where $g$ denotes the Land{\'e}-factor 
and $M$ the magnetic quantum number. Taking as an example the 
${^4}D$ term of Fe\,{\sc ii} multiplet 74 (which gives rise to
the well-known lines at $\lambda\,6147$ and at $\lambda\,6149$), 
the magnetic splitting of the $J = 3/2$ level can exceed the 
distance to the neighbouring $J = 1/2$ level for fields of about 
4.7\,T. From detailed calculations it emerges that deviations from 
simple Zeeman splitting occur much earlier: very close scrutiny of
the respective positions of the blue and red doublet components 
relative to the zero field wavelength reveals a difference of about 
20\,m{\AA} (the total splitting is 330\,m{\AA}) at only 0.7\,T (see
Fig.\,\ref{fig:6147_49_split}).

\begin{figure}
\includegraphics[width=84mm]{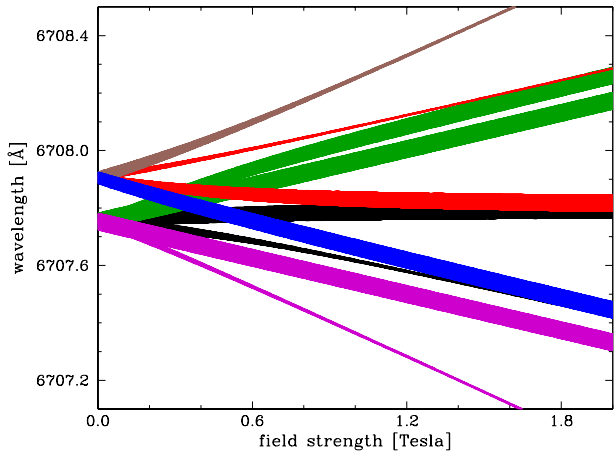}
\caption{
Wavelength splittings and relative intensities of the subcomponents of 
the Li\,{\sc i} $\lambda\,6707$ doublet as a function of magnetic 
field strength. For clarity, the respective $\pi$ components are
plotted in black and red, the $\sigma_{\rm blue}$ components in magenta
and blue, and the $\sigma_{\rm red}$ components in green and brown.}
\label{fig:li_split}
\end{figure}

The famous Li\,{\sc i}\,$\lambda\,6707$ doublet is another interesting
case that has attracted some attention over the past decades
(let us mention Engvold et al. (1970), Mathys (1991), and Leone (2007)). 
Its fine structure splitting amounts to a mere 0.34\,cm$^{-1}$ 
and can be exceeded by magnetic splitting already at field strengths 
of 0.5\,T. Again, detailed calculations show that deviations from 
simple Zeeman splitting become clearly visible somewhat earlier than
0.2\,T (Fig.\,\ref{fig:li_split}). The surprise however comes 
when one looks at the relative intensities of the 4 sub-components of 
$\lambda$\,6707.912: already at a bare 0.05\,T, the two $\pi$-components 
differ by 20\,\%, and the respective $\sigma$-components by 10\,\%! At 
the still very moderate field strength of 0.1\,T, these values rise to 
44\,\% and 20\,\%.

How many spectral lines will then be subject to the partial 
Paschen-Back effect in strongly magnetic Ap stars? This is no easy 
question to answer but a quick search in the NIST energy level 
tables already provides a substantial number of candidate terms for
some elements: there are for example 18 terms of Cr\,{\sc i} with
energies up to 52000\,cm$^{-1}$ where at least 1 pair of neighbouring 
levels are separated by less than 5\,cm$^{-1}$. It is mostly possible 
to identify the lines originating from these terms in the Kurucz
(1993b) line-list and the surprising result is that a staggering 18\,\%
or 2320 lines out of some 13000 may be affected by the partial Paschen-Back
effect in Ap stars with strong magnetic fields. The lowest ${^5}G$
term alone where the splittings between adjacent levels vary between
0.25\,cm$^{-1}$ and 2.77\,cm$^{-1}$ gives rise to 846 lines. Are there 
important lines among these? Have they been used for example for 
magnetic Doppler imaging or for abundance analyses? For the moment we
do not know; we need not be overly alarmist but the possibility cannot 
be ruled out and it is not quite clear what the consequences would be.

\begin{figure}
\includegraphics[width=84mm]{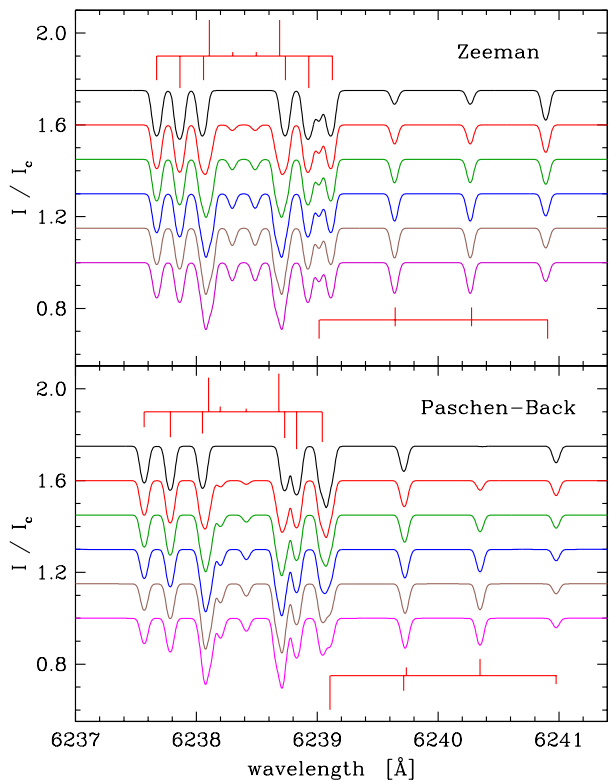}
\caption{
Stokes\,$I$ profiles for the Fe\,{\sc ii} lines at $\lambda\,6238.392$ and 
$\lambda\,6239.953$ belonging to multiplet 74, both in the Zeeman and in the 
partial Paschen-Back regime. Calculations have been carried out for a 12000\,K, 
$\log g = 4.0$ Kurucz (1993a) atmosphere, the magnetic field strength is 
2\,T, and the respective angles between field direction and line-of-sight are 
0, 30, 45, 60, 75, and 90\degr (from top to bottom). The corresponding Zeeman
and Paschen-Back patterns of the two lines are shown in the conventional form 
of vertical bars, with the $\pi$-components above and the $\sigma$-components 
below the axis.}
\label{fig:6238_I}
\end{figure}

\section{Partial Paschen-Back effect and Stokes profile modelling}
\label{sec:model}

The tool for our modelling investigations into the partial
Paschen-Back effect is CossamPaschen, a new version of the
COSSAM code developed by Stift (2000). COSSAM is a state-of-the-art
line synthesis code (Wade et al. 2001) that allows the calculation 
of full Stokes spectra either in the solar case (one specific point 
on the solar surface) or the stellar case (spatial integration of 
the profile over the visible hemisphere of the star). The code is 
characterised by LTE in a plane-parallel atmosphere, continuous 
opacities taken from Atlas9 or Atlas12 (Kurucz 1993a, 2005), full 
component by component opacity sampling (CoCoS), and by the choice 
between the Zeeman Feautrier (Alecian \& Stift 2004) and the DELO 
(Rees et al. 1989) formal polarised solvers.

In the Zeeman regime where subcomponent splittings can be assumed
to scale linearly with magnetic field strength, one just has to
store the respective anomalous Zeeman patterns for each line in 
the atomic data list. Splittings for any point on the stellar
surface are determined by simple multiplication with a 
field-dependent factor. Relative intensities are independent of 
field strength which implies that both in the solar and in the 
stellar case, storage requirements are restricted to just one set 
of Zeeman splittings and relative intensities of the Zeeman 
subcomponents.

Not so in the partial Pachen-Back regime. We have seen that 
splittings and relative subcomponent intensities go non-linearly 
with field strength. For each point on the stellar surface (and
there have to be several hundreds to a few thousands of points
depending on field geometry and on rotation) one has to determine 
the exact splittings and relative intensities for the given local 
field strength by the usual diagonalisation of the total Hamiltonian 
as outlined before. One then has the choice of synthesising 
separately all the local profiles before combining them to an 
integrated profile or to store the local Paschen-Back patterns in 
some appropriate structure and proceed first with spatial integration 
at each wavelength point. Memory requirements will be substantial, 
especially in the latter case.

Since few if any of the lines of astrophysical interest which exhibit 
the transitions to the Paschen-Back effect are entirely unblended, 
it is important to account for blending and to include the full list 
of lines appropriate to the stellar atmosphere. The final version of 
CossamPaschen thus allows the modelling of the full Stokes profiles 
of the transitions in one multiplet under the partial Paschen-Back 
regime, blended with all other transitions found in the interval in 
question and treated in the Zeeman approximation.

\begin{figure}
\includegraphics[width=84mm]{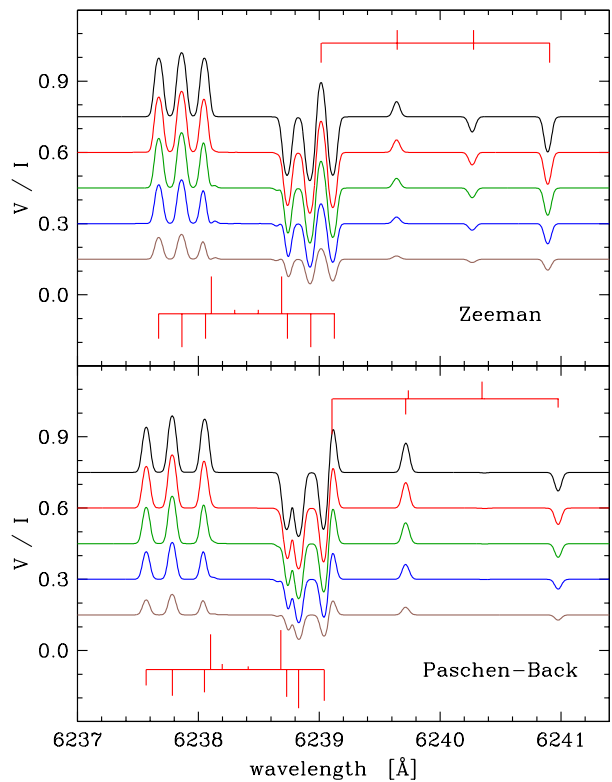}
\caption{
The same as in Fig.\,\ref{fig:6238_I}, but for Stokes\,$V$. The
case of 90\degr is omitted.}
\label{fig:6238_V}
\end{figure}

\section{Stokes profiles: Zeeman vs. Paschen-Back}

Differences between Stokes profiles calculated in the simple
Zeeman regime and profiles calculated in the partial Paschen-Back
regime can range from subtle to spectacular. Zeeman components
are shifted, get stronger or fade, symmetries are destroyed,
blends start to look quite different. The lines $\lambda$\,6238.39
and $\lambda$\,6239.95 of the Fe\,{\sc ii} multiplet 74 
beautifully illustrate these changes for a field strength of 2\,T.
The former line, a ${^4}D_{3/2} - {^4}P_{3/2}$ transition, and
the latter, a ${^4}D_{1/2} - {^4}P_{3/2}$ transition, have together
6 $\pi$ components and 5 $\sigma_{\rm blue}$ and $\sigma_{\rm red}$ 
components.

The most spectacular changes occur for a longitudinal field. The component
near 6240.4\,{\AA} disappears while at the same time the component near
6239.6\,{\AA} increases in strength (see Fig.\,\ref{fig:6238_I}). The 
outer $\sigma_{\rm blue}$ component becomes stronger too, and the blend
with the $\sigma_{\rm red}$ components of $\lambda$\,6238.39 completely 
changes shape. In Stokes $V$ we also note the conspicuous weakening of 
the $\sigma_{\rm red}$ components of  $\lambda$\,6239.95; a positive $V$ 
signal emerges to the right of the blend (Fig.\,\ref{fig:6238_V}).
Changes remain clearly visible in a transversal field, although 
they are no longer as important as in the longitudinal case. Note that 
in the Paschen-Back regime, some components hardly change intensity with
angle.

The reason for the disappearance of the component near 6240.25\,{\AA} is
easily explained: $\pi$- and $\sigma_{\rm red}$-components are almost exactly
at the same position, but the intensity of the $\sigma_{\rm red}$-component 
is now by more than a factor of 30 lower than in the Zeeman regime. The 
intensity of the red $\pi$-component grows by about a factor of 1.6. The inner 
$\sigma_{\rm blue}$-component of $\lambda$\,6239.95 displays an intensity 
increase by a factor of 2.5, the outer one by a factor of 1.8, whereas the 
blue $\pi$-component weakens.

\begin{figure}
\includegraphics[width=84mm]{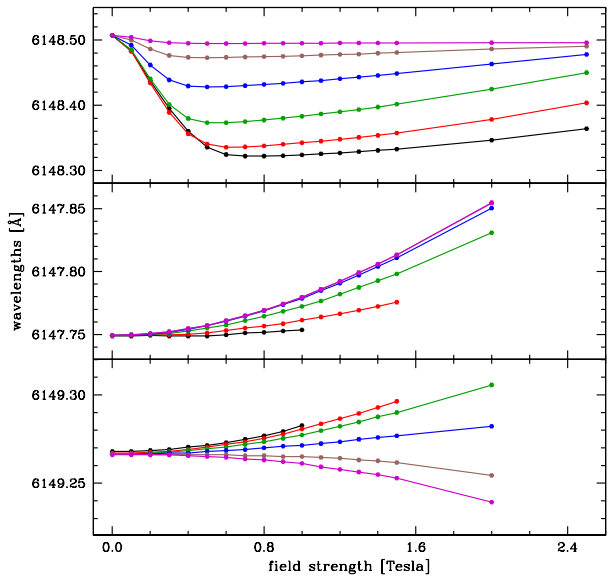}
\caption{
Wavelengths of the respective centres-of-gravity in Stokes $I$ of the 
synthesised Fe\,{\sc ii} lines at ${\lambda}\,6149$ (bottom) and at 
${\lambda}\,6147$ (middle) as a function of magnetic field strength 
and of abundance. The curves (black, red, green, blue, brown, magenta) 
correspond to decreasing abundances in steps of 0.5\,dex. Wavelengths 
of the centre-of-gravity of the 2 lines taken together are shown in
the top panel. The field is taken to be longitudinal.}
\label{fig:cog_solar_00}
\end{figure}

\section{Selected results}

There has not been in the past any systematic investigation of the 
incomplete Paschen-Back effect in the Stokes spectra of strongly 
magnetic Ap stars. Only the pioneering work by Mathys (1990) has 
provided valuable insight and raised interesting questions. We have 
tried to extend these results and we want to illustrate here some
salient points of our findings, especially concerning the 
Fe\,{\sc ii} doublet at 6149\,{\AA} and its use in the determination 
of the magnetic field modulus.

\subsection{Line shifts}
\label{sec:shifts}

The findings presented in Sec.\,\ref{sec:energy} warrant detailed 
modelling of selected lines in order to determine the wavelength 
shifts that could show up in the incomplete Paschen-Back regime. 
Indeed, the notable deviations from linear Zeeman splitting in a 
number of well-studied lines, together with the rapidly changing 
relative intensities of the subcomponents should necessarily not 
only lead to asymmetries in the line profiles but also to line
shifts of various amounts and different signs. Such shifts are
found e.g. for the Fe\,{\sc ii} line at 6149\,{\AA} in HD\,126515 
(see also Figs.\,7 and 8 of Mathys (1990) and cannot be 
modelled in a pure Zeeman regime.

For instructive purposes we have therefore chosen the Fe\,{\sc ii} 
${\lambda}\,6147$ and ${\lambda}\,6149$ lines and calculated high
resolution synthetic profiles for various field strengths up to 3\,T 
and for a number of abundances in steps of 0.5\,dex. Considering
the longitudinal and the transverse case, we obtained a grid of 228
models which cover enough of parameter space to allow a meaningful
discussion.

Based on our polarised spectral line synthesis, we note for a
longitudinal field of 2\,T and in the weak-line limit a marginal 
blue-shift (-0.027\,{\AA}) of the centre-of-gravity (COG) of the 
${\lambda}\,6149$ doublet and a clear red-shift (+0.11\,{\AA}) 
of ${\lambda}\,6147$ (see Fig.\,\ref{fig:cog_solar_00}). The COG of 
the 2 lines combined remains almost unaffected (-0.01\,{\AA}). The 
COGs are determined from the Stokes $I$ profiles. Interestingly, for 
very strong lines and 1\,T (larger field strengths lead to blending), 
the situation is partially inversed with the ${\lambda}\,6149$ doublet 
slightly (+0.015\,{\AA}) red-shifted and ${\lambda}\,6147$ essentially
unchanged. The COG of the 2 combined lines however is now blue-shifted
by a remarkable -0.18\,{\AA}).

\begin{figure}
\includegraphics[width=84mm]{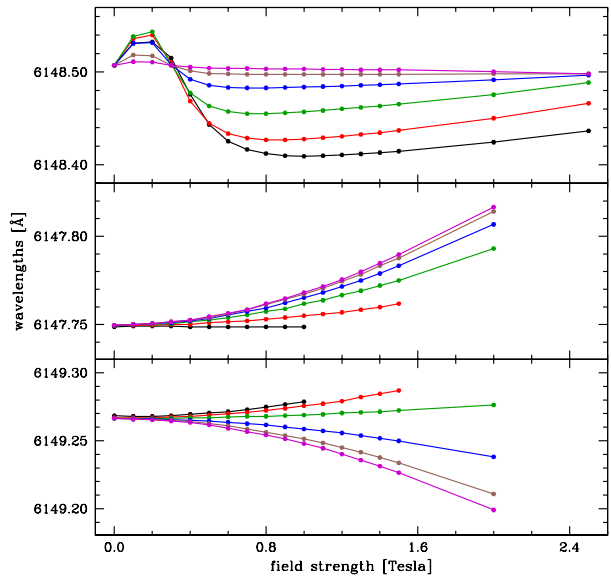}
\caption{
The same as in Fig.\,\ref{fig:cog_solar_00} but for a transverse field.}
\label{fig:cog_solar_90}
\end{figure}

In the case of a transverse 2\,T field, the weak-line limit leads to a 
noticeable (-0.07\,{\AA}) blue-shift of ${\lambda}\,6149$ and a red-shift 
of exactly the same size (+0.07\,{\AA}) of ${\lambda}\,6147$ (see 
Fig.\,\ref{fig:cog_solar_90}). Again, the COG of the 2 lines combined
remains almost unaffected (-0.01\,{\AA}). Very strong lines and a 1\,T field 
yield a marginally (+0.01\,{\AA}) red-shifted ${\lambda}\,6149$ doublet and 
${\lambda}\,6147$ unchanged. The blue-shift of the COG of the 2 lines combined 
is reduced to -0.10\,{\AA}) but surprisingly we now also encounter a red-shifts 
of up to 0.025\,{\AA} for fields of less than 0.3\,T.

Decidedly, the behaviour of the various lines of multiplet 74 of
Fe\,{\sc ii} in the incomplete Paschen-Back regime cannot easily be 
predicted. Line shifts, intensities and asymmetries vary with magnetic 
field strength, field inclination and total line opacity in the 
multiplet. The same holds true for the numerous other lines in the
solar spectrum and in stellar spectra which show the incomplete 
Paschen-Back effect.

\begin{figure}
\includegraphics[width=84mm]{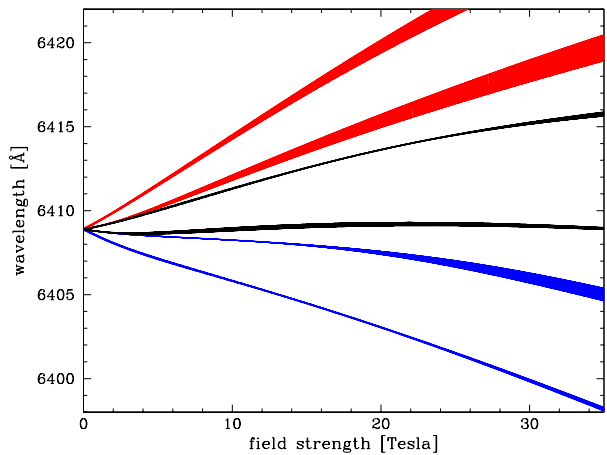}
\caption{
Splittings and relative intensities of the subcomponents of the
``ghost line'' at $\lambda$\,6408.9 (belonging to multiplet 74 of
Fe \,{\sc ii}). The respective $\sigma_{\rm blue}$, $\pi$ and
$\sigma_{\rm red}$ components are plotted in different colours.
}
\label{fig:Fig_6408}
\end{figure}

\subsection{Ghost components}
\label{sec:ghosts}

As pointed out in Condon \& Shortley (1935) and in more detail in  
Landi Degl'Innocenti \& Landolfi (2004), lines appear in the 
incomplete Paschen-Back regime which are normally forbidden under the 
selection rules for electric dipole transitions. This has to be
ascribed to the fact that $J$ is no longer a good quantum number and
there occurs J-mixing of the various levels belonging to a particular
term. These lines are sometimes referred to as ``forbidden lines'', 
but we prefer here to call them ``ghost components'' or ``ghost lines'' 
in order to avoid confusion with the well known forbidden lines in 
astrophysics which arise from metastable levels.

Take multiplet 74 of Fe\,{\sc ii} as an example. Without a
magnetic field we have 8 transitions at $\lambda\,6456.383,$ 
$6416.919,$ $6407.251,$ $6247.557,$ $6239.953,$ $6238.392,$ 
$6147.741,$ and $6149.258$, but J-mixing leads to an additional 
4 transitions at $\lambda\,6408.897,$ $6284.959,$ $6192.961,$ 
and $6156.642$. At a field strength of 4\,T, the total strength 
of the $\lambda 6408.897$ ``ghost line'' reaches 15\% of the 
strength of the nearby allowed $\lambda 6407.251$ line. Generally, 
the ``ghost lines'' are extremely weak in fields of the order of 
0.1\,T but their strength can become comparable to that of 
many allowed components once field strengths of 10\,T are 
exceeded. 

This led Mathys (1990) to speculate that some ghost lines 
originating in strong multiplets of abundant elements could 
possibly become observable in selected Ap stars whose atmospheres 
are permeated by fields in the range of $3 - 5$\,T in some parts. 
We are rather sceptical as for the observability for reasons 
obvious from Fig.\,\ref{fig:Fig_6408}: ``ghost lines'' can show 
up only in fairly strong fields so that the observable profiles 
will invariably be severely diluted unless the star is not only 
endowed with a reasonably extended strong-field region but also 
exhibits only weak field gradients.

\subsection{The Fe\,{\sc ii} doublet at 6149\,{\AA} and the 
         magnetic field modulus}
\label{seq:doublet}

Originally, the main rationale for our investigation was to establish 
whether or not measurements of the magnetic field modulus $H_{s}$ 
from the splitting of the Fe\,{\sc ii} $\lambda\,6149$ Zeeman 
doublet could be affected by the incomplete Paschen-Back regime and 
possibly lead to incorrect results when interpreted in terms of
standard Zeeman splitting. 

\begin{figure}
\includegraphics[width=84mm]{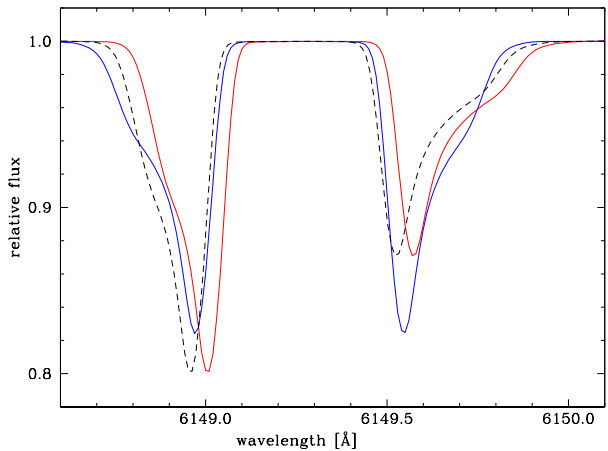}
\caption{
A comparison between the mean profiles from 9 different Oblique Rotator
models as detailed in the text, calculated in the partial Paschen-Back 
regime (red) and in the Zeeman regime (blue). The wavelength shift
PB minus Zeeman is -0.048\,{\AA}. The dashed profile (black) represents 
the PB-profile shifted by this amount. allowing a better comparison of
the respective {\em shapes} of the profiles.}
\label{fig:fe_mean}
\end{figure}

To this end we have calculated a grid of profiles for 9 Oblique Rotator 
models with various inclinations and excentricities, all with a field 
modulus of $H_{s} = 1.45$\,T, but with dipole strengths ranging from 
0.86 to 1.40\,T (in a centred dipole model, the polar field strengths
would be twice these values). Means were taken over the wavelengths of 
the 9 respective centres-of-gravity, blue and red component
separately, both for the incomplete Paschen-Back and the Zeeman
regime. PB-profiles are found to be {\em always} red-shifted by about 
0.05\,{\AA} relative to the Zeeman-profiles, but there is no change in
the splittings (see Table\,\ref{tab:COG_positions}); the value of 
$H_{s} = 1.442$\,T in the former regime is nearly the same as 
$H_{s} = 1.457$\,T in the latter. PB-profiles can become dramatically 
asymmetric: a deep and relatively narrow blue component stands in
sharp contrast  to a shallower red component characterised by a strong 
and extended redward wing (Fig.\,\ref{fig:fe_mean}).

These models (and all other models calculated on a random basis) have 
not revealed any undesirable behaviour of the doublet as far as the
determination of $H_s$ is concerned, although, as discussed above, 
sizeable wavelength shifts are encountered. With all the necessary
caution, we deem it highly probable that these findings hold true for 
all magnetic geometries.

Let us note that the scatter about the mean COG wavelengths in 
Table\,\ref{tab:COG_positions} is extremely small; PB-profiles exhibit 
somewhat larger differences between each other than the
Zeeman-profiles. This suggests that at least some of the lines 
formed in the incomplete Paschen-Back regime can provide increased 
diagnostic content for magnetic field mapping. Inclusion and correct
treatment of these lines is expected to increase the reliability of 
magnetic maps.

\subsection{Doppler mapping}
\label{seq:doppler}

\begin{figure}
\includegraphics[width=84mm]{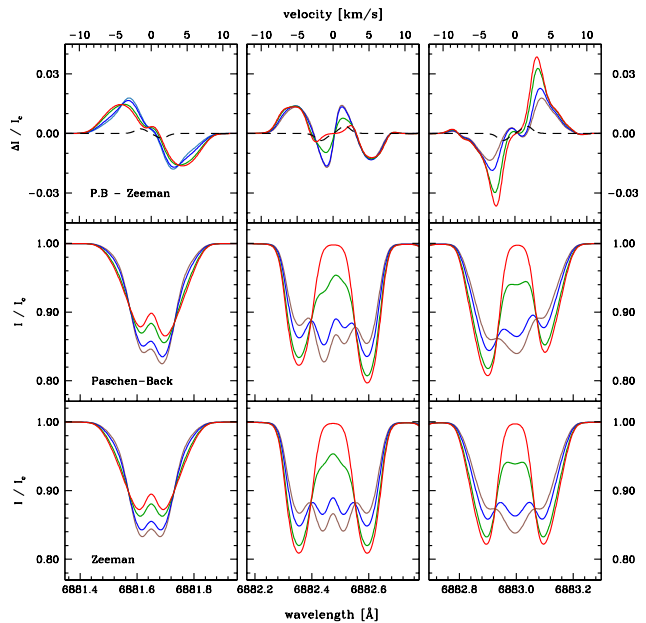}
\caption{Intensity profiles of the lines of multiplet 222 of Cr\,{\sc i} in the 
spectral interval $6881 - 6884$\,{\AA} for a 7500\,K, $\log g = 4.0$ Kurucz
(1993a) model. The magnetic field strength is 0.25\,T, the respective 
angles between the field direction and the line of sight are 0, 30, 60, 
and 90\degr (red, green, blue, brown). Note the perfect symmetry of the 
components in the Zeeman regime (bottom) and its deformation by the partial 
Paschen-Back effect (middle). The relative difference between the two profiles 
reaches about 20\% (top). Since the theoretical line strengths and positions do 
not perfectly match the Kurucz line data, very small differences occur even for 
zero field strength (dashed black line in the top panel). For clarity we have 
indicated both the wavelength and the velocity scale.}
\label{fig:diff}
\end{figure}

A few words of caution are appropriate for the aficionados of (magnetic) Doppler 
imaging. In fact, if one looks at the list of 12 Cr\,{\sc i} lines used by 
Kochukhov et al. (2004) for the mapping of HR\,3831, the lines at 6881.65, 
6882.48, and 6883.00\,{\AA}, all belonging to multiplet 222, are definitely in 
the partial Paschen-Back regime, given the
narrow wavelength spread of these 3 lines. The quite uncertain oblique 
rotator model with a centred dipole proposed by these authors -- detailed
analyses based on full Stokes profiles usually find non-negligible quadrupole 
contributions -- predicts a polar field strength of 0.25\,T. We calculated local 
intensity profiles for this field and various angles between field vector and
line of sight. The respective Zeeman and PB intensity profiles differ by up to 
20\% as can be seen in Fig.\,\ref{fig:diff}. 

\begin{table}
\caption{Mean wavelength positions in the Zeeman and in the partial 
Paschen-Back regime of the centres-of-gravity of the blue and red components 
respectively of the $\lambda$\,6149 Zeeman doublet for different oblique
rotator geometries but a constant field modulus of $H_{s} = 1.45\,T$. 
Wavelengths are given in {\AA}, the scatter $\sigma$ in m{\AA}.}
\label{tab:COG_positions}
\begin{tabular}{l|r|r|r|r|r|r}
 &  \multicolumn{1}{c}{${\lambda\,}^{COG}_{blue}$} & \multicolumn{1}{|c}{$\sigma$} 
 &  \multicolumn{1}{c}{${\lambda\,}^{COG}_{red}$}  & \multicolumn{1}{|c}{$\sigma$}
 &  \multicolumn{1}{c}{${\Delta\lambda\,}^{COG}$}  & \multicolumn{1}{|c}{$\sigma$} \\
\hline
Zeeman & 6148.916 & 0.4 & 6149.602 & 0.3 & 0.686 & 0.5\\
P.-B.  & 6148.968 & 0.5 & 6149.647 & 0.5 & 0.679 & 0.5
\end{tabular}\\
\end{table}

Translated to velocity space, the use of Zeeman profiles instead of the correct
PB profiles will thus result in spurious over- and under-abundances at relative 
velocities ranging from $\pm 3$\,km\,s$^{-1}$ to $\pm 6$\,km\,s$^{-1}$ in these
3 lines. Note that the differences between PB and Zeeman profiles hardly depend
on the angle in the case of $\lambda\,6881.65$, whereas a strong dependency is
found for $\lambda\,6883.00$. In the latter line, differences are largest for a
longitudinal field but just the opposite is true for $\lambda\,6882.48$. Looking
at the sign of the differences, one finds a global blue shift of the
$\lambda\,6883.00$ PB profile and a global red shift of the other 2 lines.

Given this complex behaviour, it is difficult if not impossible to predict to
what degree the Cr abundance maps derived by Kochukhov et al. (2004) may have 
been affected by the simplifying assumption of normal Zeeman splitting for 
these 3 lines. Close scrutiny of their Fig.\,6 reveals that whereas 
$\lambda\,6881.65$ and $\lambda\,6882.48$ are reasonably well reproduced by 
the chromium Doppler map shown in their Fig.\,8, there are significant deviations 
from the observed profiles in the vicinity of $\lambda\,6883.00$ near phases 
0.30 and 0.78 (when the magnetic equator dominates the visible hemisphere and 
the poles are near the limb). Given the special behaviour of $\lambda\,6883.00$,
can we attribute these differences to pure chance or is there some
responsibility of the PB effect? The question remains open and further detailed 
investigations are necessary to assess the true importance of this potential 
problem.

\section{Conclusions}
\label{seq:conclusion}

High resolution spectral observations of magnetic Ap stars make it clear that
the Paschen-Back effect cannot be neglected in the interpretation and modelling
of a number of spectral lines, whether it be the Li\,{\sc i} $\lambda\,6707$ 
resonance line or the Fe\,{\sc ii}\,$\lambda\,6149$ Zeeman doublet. Because of
its simple structure, the latter is particularly useful for the determination of 
the magnetic field modulus $H_s$. We have calculated in detail the splittings of 
the spectroscopic terms involved in these transitions and the relative
intensities of the subcomponents and found, somewhat to our surprise, that some
lines can enter the incomplete Paschen-Back regime already at field strengths
of a bare 0.05\,T as is the case for the lithium resonance lines. Looking
further to other chemical elements, we have discovered that in extreme cases
(Cr\,{\sc i} is an example), up to an estimated 18\% of the lines may exhibit
the transition to the partial Paschen-Back regime in Ap stars with strong 
magnetic fields of the order of 2\,T.

The development of CossamPaschen, a modified version of COSSAM -- the polarised 
line synthesis code established by Stift (2000) -- has made it possible
over the last few years to explore the effect of the partial Paschen-Back
regime on Stokes spectra. CossamPaschen allows the detailed and realistic 
modelling of a PB multiplet, blended with lines which are treated in the Zeeman 
approximation. First ever detailed Stokes profiles in realistic stellar 
atmospheres have been presented at the CP\#AP Workshop in Vienna (Stift 2007) 
and provide valuable insight into the sometimes spectacular changes of selected 
spectral lines. These calculations show that the centres-of-gravity of the
Fe\,{\sc ii}\,$\lambda\,6149$ line and of its sister line at $\lambda\,6147$ are 
shifted in magnetic fields, either towards the red or towards the blue,
depending on line strength, magnetic field strength, and field direction. A
comparison between the observations plotted in Fig.\,\ref{fig:shift} and the
theoretical results displayed in Figs.\,\ref{fig:6238_I} and \,\ref{fig:6238_V}
shows gratifying agreement.

Another interesting finding is the confirmation of the existence of normally
forbidden ``ghost components'' in strong magnetic fields. The intensity of these
components is a very strong function of magnetic field strength, and in Ap stars 
some of them may actually significantly contribute to the {\em local} spectra in
places with very strong magnetic fields (of the order of 4\,T). However, the
strong dependency of intensity and position on field strength will probably lead 
to the {\em global}  signature of the ``ghost components'' begin washed out.

An important result of our investigation is the confirmation that $H_s$
measurements obtained from the splitting of the Fe\,{\sc ii}\,$\lambda\,6149$
line and interpreted in terms of classical Zeeman splitting do not need to
be corrected for the PB effect. For all practical purposes, even in the
strongest fields encountered in Ap stars, the PB splitting is not different from
the Zeeman splitting. As for maps derived by means of (magnetic) Doppler
mapping, we cannot provide such a reassuring confirmation and there is a real
danger that the inclusion of lines subject to the partial PB regime may lead to
spurious results. What can be a serious nuisance may also provide improved
diagnostic capabilities and so we are looking forward to improved imaging codes 
incorporating the full treatment of the partial Paschen-Back effect.

\section*{Acknowledgements} 
MJS acknowledges support by the {\sf\em Austrian Science Fund (FWF)}, 
project P16003-N05 ``Radiation driven diffusion in magnetic stellar 
atmospheres''.

\bsp

{}

\label{lastpage}

\end{document}